# MATH WORLD: A GAME-BASED 3D VIRTUAL LEARNING ENVIRONMENT (3D VLE) FOR SECOND GRADERS

Jean Maitem[1] , Rosmina Joy Cabauatan[2], Lorena Rabago[3], Bartolome Tanguilig III[4]

[1]Department of Information Technology Education, Technological Institute of the Philippines, Quezon City, Philippines
jeanmaitem@gmail.com
[2]Department of Information Technology Education, Technological Institute of the Philippines, Quezon City, Philippines
rjmcabauatan@yahoo.com
[3] Department of Information Technology Education, Technological Institute of the Philippines, Quezon City, Philippines
lwr823@yahoo.com
[4] Department of Information Technology Education, Technological Institute of the Philippines, Quezon City, Philippines
bttanguilig_3@yahoo.com

## *ABSTRACT*

*This paper intends to introduce a game-based 3D Virtual Learning Environment (VLE) to second graders. The impetus arose from the need to make learning in mathematics more effective and interesting through multimedia. Applied in a game, the basic mathematical operations such as addition, subtraction, multiplication, and division are expected to performed by learners as they represent themselves as avatars while they immerse in a quest of digital objects in the VLE called Math World. Educational attributes such as mentality change, emotional fulfillment, knowledge enhancement, thinking skills development, and bodily coordination are evaluated to ensure learning effectiveness. Also, game playability measured in terms of game plays, story, mechanics and interface usability are examined for its educative design. With an aggregate of these enhanced indices, results attest that objectives were met while making mathematics an interesting, motivating and enjoyable subject, hence VLE a significant tool to complement the conventional approaches of teaching.*

## 1. INTRODUCTION

In recent years, elementary educators of different levels have experienced a rapidly increasing demand for flexibility in the way how teaching and learning in mathematics are facilitated in order to make it more effective and motivating [5][6] since most learners considered it to be the most difficult among other subjects. One of the key implications of this demand is the need for innovation in the design of learning resources as alternative to face-to-face classes [4]. Educators started the use of information and communication technologies like email, social networking sites, learning management systems, instant messaging services, wikis, blogs and Voice over Internet Protocols (VoIP). However, to date, a relatively young educational technology called 3 Dimensional (3D) Virtual Learning Environment (VLE) is gaining popularity as it offers potentials and supplements to pedagogical routines more than the typical approaches. When applied to mathematics, this allows the learner to manipulate objects within the environment in order to develop a much greater level of understanding [5].

The environment is created entirely from a computer database consisting of objects modeled by computer-aided design (CAD) software. These objects are programmed to behave in certain ways as the learner interacts with them [21]. Different approaches in teaching are also provided necessary for information sharing, in which learners guide avatars through the virtual world. Avatars are graphical representations of learners or characters in games. Typically, teaching and learning programs make use of a multi-learner or single-learner virtual environment to immerse in educational tasks. The task may be termed a quest, mission or challenge, depending on a scenario [17]. Learners can move around to virtual places while performing these tasks. In pursuance, interaction with digital objects is required while acting as avatars [17].

Along with the educative potentials of VLEs are the great incentives of games as educational tools. From the viewpoint of game theory, learners could practice critical thinking for decision-making, and help construct their own concepts and knowledge while they accomplish the objectives of the game [12]. With multimedia contents of games such as videos, images, text, and audio, the learner gains knowledge and skills by solving problems while playing. However with multimedia design for education, it should combine interactive design and motivational content with the most effective principles of technologically mediated learning. [12]. Pedagogical approaches that were found effective include learning by doing, learning from mistakes, goal-oriented learning, and constructivist learning. Major genres of games include action games, adventure games, fighting games, role playing games, simulation games, sports games and strategy games. These have different learning effects, which are achieved due to their structural elements such as rules, goals and objectives, outcomes and feedback, conflict competition, challenge and opposition, interaction and representation [18].

For this study, a game-based VLE called Math World was developed for second graders. Lessons on mathematical operations such as addition, subtraction, multiplication, and division are designed along with the objective of gauging its effectiveness towards learning in the form of adventures and simulations. As a game, introduction of new knowledge, fixing of previous knowledge, skills, and discovery of new concepts are integrated on all areas of the game in incremental way allowing transition from the basic operations to more advanced topics. Contents are based from the prescribed Department of Education (DepEd) mathematics textbook for second graders.

The rest of the paper is organized as follows: review of related literature, gaming model, and system architecture. Other sections include evaluation and result, conclusion and future works.

## 2. RELATED LITERATURE

VLE has begun to be used on the internet as learning resource. By nature, its components are accessed from remote locations[1] and intended for learners to inhabit, and socially interact [3][1]. 3D environment has been widely used to complement the typical virtual environment in which learners take the form of avatars in order to be graphically visible to others[3]. Few examples of virtual social interactions include instant messaging, discussion boards, emails, blogs, and podcasts [1]. Added with immersive contents, a 3D VLE allows learner to explore and learn at his own pace and time. However, a teacher needs to scaffold a lesson to ensure that learners are able to follow the lessons. [17].

3D VLE is modeled using 3D vector geometry. The learner's view of the environment is rendered dynamically according to their current position in 3D space, that is, the learner has the ability to move freely through the environment and their view is updated as they move. At least some of the objects within the environment respond to learner action, for example, doors might open when approached and information may be displayed when an object is selected with a mouse. Some environments include 3D audio, that is, audio that appears to be emitted from a source at a particular location within the environment. The volume of sound played from each speaker depends on the position and orientation of the learner within the environment [17].

Many virtual environments have been developed for educational purposes. Baxter & Amory [2] has developed an educational adventure game, Zadarh which aims to address learning misconceptions in biology in the context of the appropriate application of theories of learning. Linden Lab developed Second Life, an open source software online virtual world which allows the users to build and program virtual objects. Trindade, et al [23] developed a virtual environment of water molecules for learning and teaching science. Yair, et.al [22] research supports constructivist approach to learning in which learners learn by doing rather than reading. Hong, et al [14] developed an assessment tool to examine the educational values of digital games. Supplementing[14] is Kiili's[15] playability properties integrated in game interfaces. Among these researches, a combination of the indices used to measure the educational qualities of a game developed by Hong, et al and the interface game usability measures of Kiili are found significantly relative to the objectives of this paper, hence enhanced and utilized to measure VLE's learning effectiveness, interestingness, and motivational and enjoyment attributes.

## 3. GAMING MODEL

Figure 1 presents the model used to develop Math world. The process starts at the center of the spiral with ideas as to how the objectives are met. It proceeds outward clockwise through each of the phases starting from definition, design, prototype, playtest, and feedback. The next section provides the details as to how these phases are executed towards the realization of the learning environment.

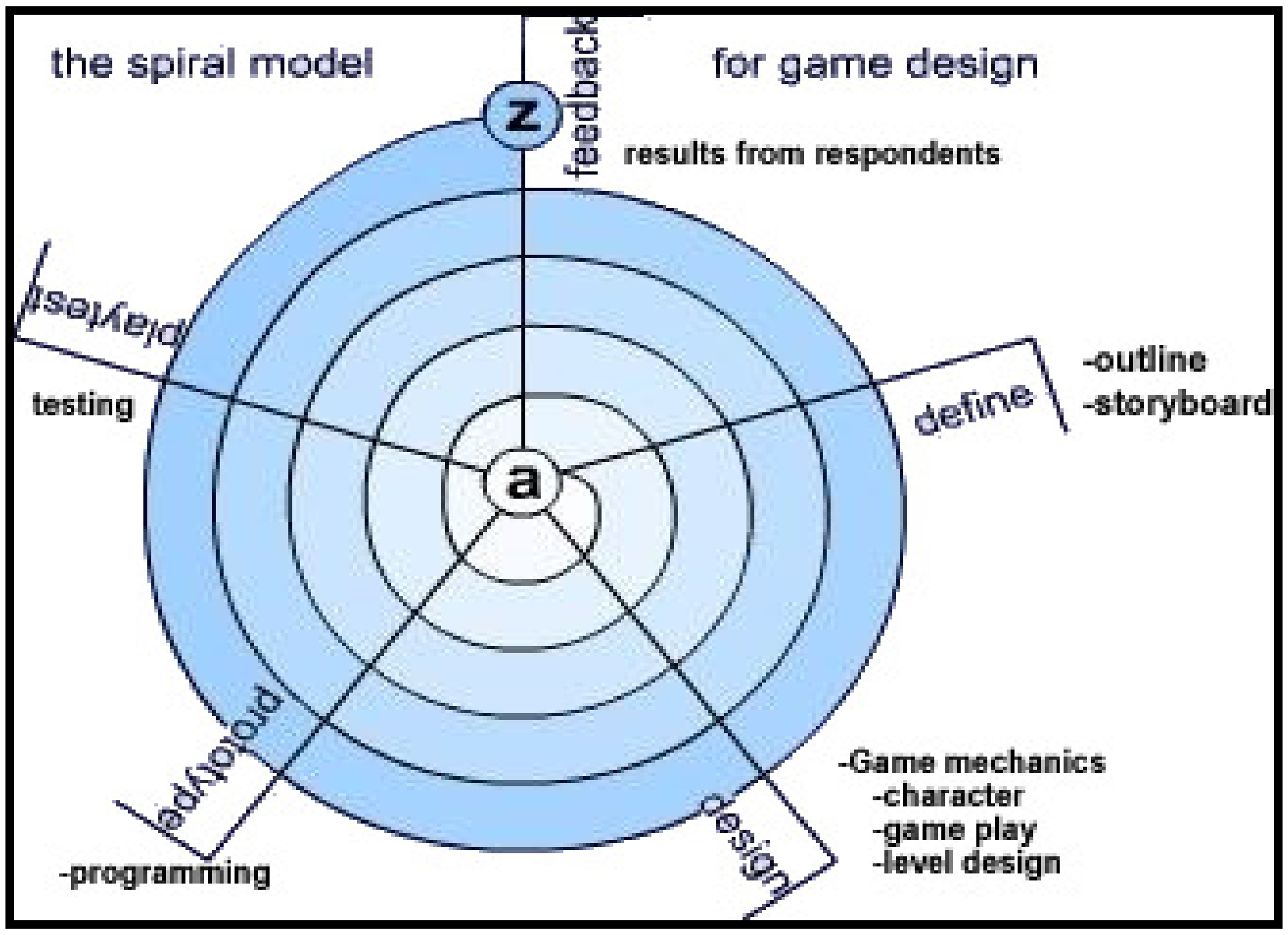

Figure 1. Spiral Model for Game Development

# 4. SYSTEM ARCHITECTURE

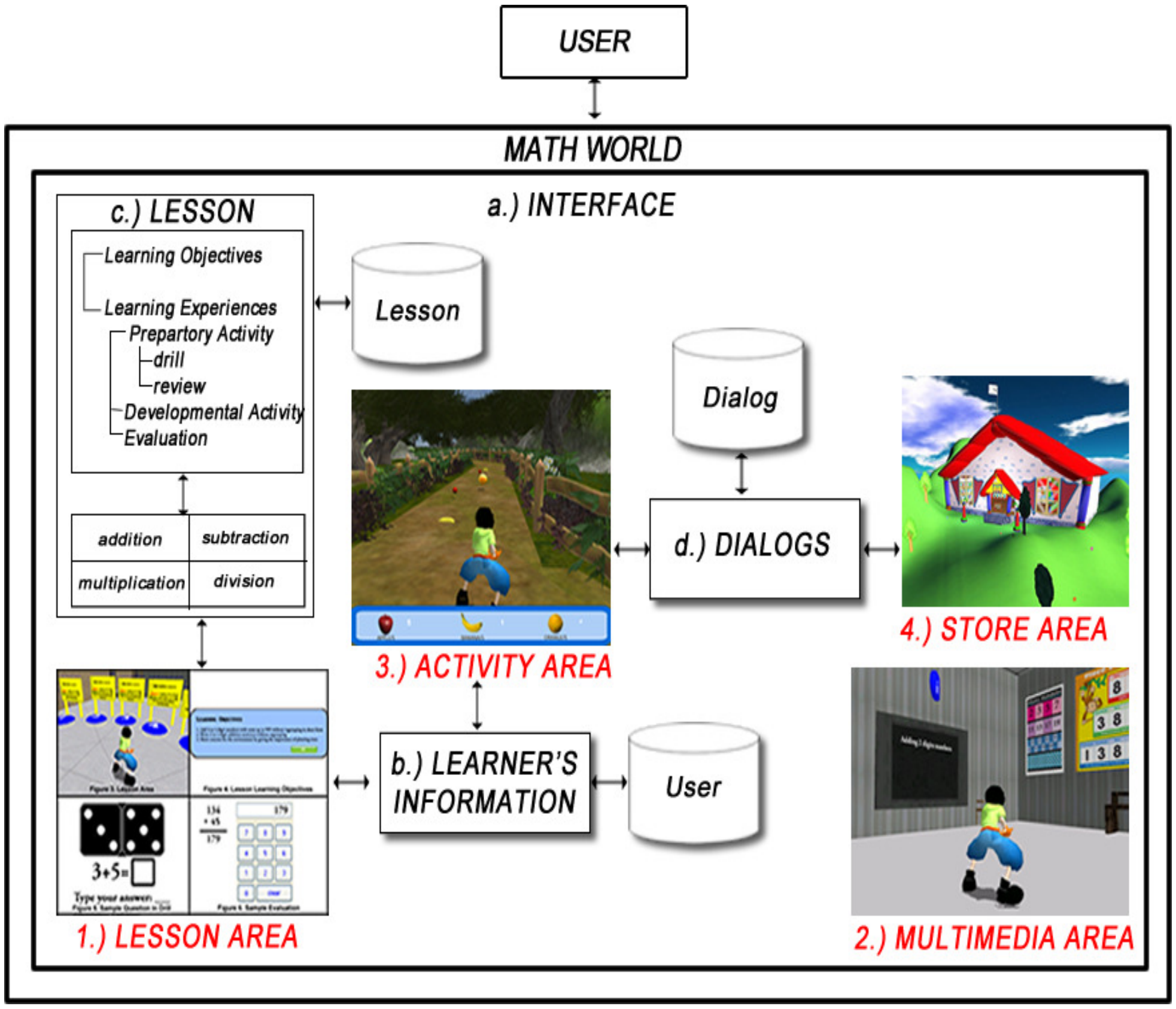

Figure 2. Math World Virtual Learning Environment

Figure 2 outlines the VLE's storyboard. It has four main modules, a) interface, b) learner's information, c) lessons, and d) dialogs. The interface guides the learner to move around the environment. The learner's information stores the profile, grades and lessons taken by the learner. Lessons contain the topics to be undertaken by the learner. Dialogs deliver strings or characters for the communication between the learner and the math world.

Math world encourages the learners to practice and develop their analytical and problem solving skills in mathematics by accepting tasks or mission in the form of adventure, quiz and games. Different areas namely 1) lesson, 2) multimedia, 3) activity and 4) store areas are explored as tasks are performed. Distinctively, when the learner runs the game, the title and menus appear on the screen. The game starts from registration and when pieces of information supplied in a form are provided, the learner goes to any areas to perform a task. Menu for volume, music and video screen resolution are provided to let the learner customize the game interfaces.

Math world lessons include addition, subtraction, multiplication and division. Table 1 exhibits the lessons and their respective topics along with the activities to be performed by learners in order to meet the objectives. Aligning these attributes is significant in measuring the effectiveness of the game.

Table 1. Lessons Structure

| Lesson | Topics | Objectives | Learning Activities | | |
|---|---|---|---|---|---|
| | | | Preparatory Activities | Developmental Activities | Evaluation |
| **Addition** | Adding 2-to-3-Digit Numbers with Sums up to 999:<br><br>-without Regrouping<br>-with Regrouping<br>-with Zero in any of the Addends without Regrouping<br>-with Zero in any of the Addends with Regrouping<br><br>Analyzing Word Problems | Solve 2-to-3 digit numbers with sums up to 999<br><br>Analyze and solve worded problems in addition | 1.Drill<br>Basic addition using a domino cards<br>2. Review<br>Adding numbers using show me board | -Solving 2-to-3 digit numbers<br>-Solving of puzzles | Quiz |
| **Subtraction** | Subtracting 2-to3-Digit Numbers<br>-without Regrouping<br>-with Regrouping in the Tens Place<br>-with Regrouping in the Hundreds Place and with Zero Difficulty<br>-with Regrouping in the Tens and Hundreds Place and with Zero Difficulty<br><br>Analyzing Word Problems | Solve and subtract 2-to-3-digit numbers with regrouping in tens place and hundreds<br><br>Analyze and solve worded problems in subtraction | 1.Drill<br>Exercises in subtracting numbers<br>2. Review<br>Subtracting numbers using show me board | -Subtracting 2-to-3 digits numbers<br>-Game: Fruit picking | Quiz |
| **Multiplication** | Multiplication as Repeated Addition using Sets<br><br>Multiplication as Repeated Addition using Number Line | Show the relationship of putting together 2 or more sets with the same number of elements to repeated addition<br><br>Show multiplication of whole numbers as repeated addition on the number line or array etc | 1.Drill<br>Addition facts in flash cards<br>2. Review<br>Solve a problem in addition<br><br>1.Drill<br>Bring Me<br>2. Review<br>Selecting correct number on the blanks | -Explaining that repeated addition can express in a shorter way through multiplication<br><br>Showing multiplication of whole numbers as repeated addition on the number line or array etc. | Quiz |
| | Identifying Parts of a Multiplication Sentence | Identify the parts of a multiplication sentence | 1.Drill<br>Skip counting by 3s,4s, 5s<br>2. Review<br>Using show-me-board | Identifying the parts of a multiplication sentence | |
| | Showing that Zero Multiplied by a Number is Zero | Show that zero multiplied by a number is zero and multiply 1-to-2 digit number with products up to 81 | 1.Drill<br>-Multiply numbers | Showing that zero multiplied by a number is zero | |
| | Multiplying 1-to-2 Digit Numbers with Products up to 81 | Analyze word problems involving multiplication of whole numbers including money | 1. Drill<br>Multiplication Facts<br>2. Review<br>-Multiply numbers | Showing a number multiplied by a 1-to-2 digit number with products up to 81 | |
| | Analyzing Word Problems | | | Analyzing word problems involving multiplication of whole numbers including money | |
| **Division** | Parts of a Division Sentence | Identify the parts of a division sentence namely dividend, divisor and quotient | 1. Drill<br>Drill on the basic division facts<br>2. Review<br>Subtract a number | Identifying the parts of a division sentence namely dividend, divisor and quotient | Quiz |
| | Illustrating Division<br>-by Partition<br>-by Distribution<br>-as Repeated Subtraction<br>-as Inverse of Multiplication | Illustrate division by partition, distribution, as repeated subtraction and inverse of multiplication | 1. Drill<br>-Fill in the blanks<br>2. Review<br>-Identify the parts of a division sentence | Illustrating division by partition, distribution, as repeated subtraction and inverse of multiplication | |
| | Dividing 1-to-2Digit Numbers by 1-Digit Number<br>Dividing 2-Digit Numbers by 1-Digit Number with Dividends through 81 | Divide numbers | 1.Drill<br>-Basic multiplication facts<br>2.Review<br>-Find the multiplication sentence that will match the division sentence | Dividing numbers | |
| | Analyzing Word Problems | Analyze word problems involving division | 1. Drill Multiplication Facts<br>2. Review<br>-Multiply numbers | Analyzing word problems involving division | |

Utilizing the educational values of a game in the research of Hong et al [14], such as mentality change, emotional fulfillment, knowledge enhancement, thinking skills development, interpersonal skills, and bodily coordination would best evaluate Math world as to the qualities of an educational game. These indices are briefly described below with emphasis of its actual use to the Math world.

Mentality change is attributed to promotion of adventure, evaluation of trade-offs and awareness of efficiency. Adventure is promoted as lesson levels increase, in which learner accept more tasks when seeking to obtain more tickets for rewards of performing activities. Along with this are trade-offs of each decisions made as the learner compels to perform the activities with time limit. Likewise, more rewards are gained when activities demands more tasks to be performed.

Emotional fulfillment is satisfied as learners experience interactive gaming process. In similar manner, as the player is required to carry out any mathematical operations in a time-sensitive nature, full attention is needed in attempt to perform more tasks.

Knowledge acquisition, reinforcement and enhancement are realized as the learner performs the tasks incrementally beginning from addition to division stage. Learners are likewise motivated to take the rewards, and as activities level up.

Thinking skills development becomes more intense as tasks require previously acquired knowledge to be applied. Also, sensory stimuli such as pictures, sounds and words encourage the learner to apply their observations and perception skills. For instance, in the multimedia area, the learner can watch video about tutorials in mathematics. Likewise, the learner can view and read posters on the wall such as multiplication table, shapes and odd/oven numbers.

When playing, the learner uses a mouse to select and link answers of questions as quickly as possible. These help the learner become more dexterous by developing hand-eye coordination skills. In the activity area, for example, a learner can play a game about a farmer (non-player character) asking for help to pick fruits. The learner goes in a farm upon accepting the task given by the farmer. However, to go inside the activity area, learner must have tickets that can be acquired in the lessons area.

Motivation is significant in game-based learning. This is the desire for change that the learner is driven by situations. Motivation to learn can be intrinsic and extrinsic.Extrinsic motivation is an involvement as a means to an end. Intrinsic motivation is an involvement for its own sake. [9]. Motivation in math world is in the form of rewards in which learners are motivated by activities or experiences that present a challenge, giving controls to decisions and evokes curiosity as the environment is explored in quest for seeking more rewards in the form of tickets. These tickets are exchanged in the store area. The learner can swap the tickets for some items such as coloring sheets.

Another potential approach is to have comparisons of assessment scores from the different game levels performed by the player. This could positively attest that learning is successfully attained through the scores in the different game plays.

As claimed by Prensky [18], the effectiveness of digital games also lay in their design. If the game designers could incorporate educational values, learning is enhanced. In [8][10], Driskell et al and Frasca state that games require continuous practice to improve accuracy and better memory and educational values in games are measured by the game objectives.

Argued by Squire [20], many game development researchers are not very interested in examining whether games offer content that may be relevant to educational values; they usually have narrow focus on game content, and skills and attitudes of learners. Opposed in this paper are these three components equally significant in measuring the educative implications of games.

Educational values of a game is not solely measured in terms of learning effectiveness[14] but also on playability and engagement[15]. These are attributed to game designs and are based from the concept of heuristic, a design guideline which serve as a useful evaluation tool for product design. When applied to measure software quality, this is used to evaluate the usability of interfaces with goals to make software easy to learn, use and master but opposed to design goals for games, usually characterized as easy to learn but difficult to master. Design is not just its physical appearance but an aggregate of game play, game story, game mechanics, and game usability. Game play are the set of problems and challenges a player must face to win a game. Game story includes all plot and character development. Game mechanics involve programming that provides the structure by which units interact with the environment and game usability addresses the interface and encompasses the elements the user utilizes to interact with the game(i.e mouse, keyboard, or controller).

The physical designs of the Math world were made colorful and creative since learners are second graders. The navigation and controls were made simple and easy for learners to easily adapt to the environment. These were achieved through 3D Game Studio, Adobe Fireworks and Adobe Photoshop.

# 5. EVALUATION AND RESULTS

In evaluating the educational effectiveness of Math World, a 5-point Likert-scaled questionnaire was designed to enable second graders to self-report on what they felt and learned from undertaking the activities of the game. This provides a meaningful level of discrimination without forcing the learner to have an opinion. The validity of the tool was examined by teachers handling the subject in order for second graders to understand the questions asked. To ensure the reliability of results, evaluation was administered to 40 incoming second graders without prior knowledge on the covered mathematical operations to be taught in Math World. Shown in Table 2 is the result of the evaluation.

**Table 2. Assessment on the Effectiveness of Math World**

| Attribute | Description | Mean | | Interpretation |
|---|---|---|---|---|
| Mentality Change | The learner can accept more tasks to obtain more tickets in exchange of rewards.<br>Activities are performed with time limit | **3.53** | Agree | Effective |
| Emotional Fulfillment | The learner is required to carry out any mathematical operations in a time-sensitive nature<br>Full attention is needed in attempt to perform more tasks. | **3.40** | Agree | Effective |
| Knowledge Enhancement | The learner performs the tasks incrementally beginning from addition to division stage.<br>Learners are motivated to take rewards as activities level up | **3.38** | Agree | Effective |
| Thinking Skill Development | Tasks require previously acquired knowledge to be applied.<br>The use of pictures, sounds and words encourage the learner to observe and apply perception skills. | **3.43** | Agree | Effective |
| Bodily Coordination | The learner uses a mouse to select and link answers of questions as quickly as possible | **4.00** | Agree | Effective |
| **Weighted Mean** | | **3.56** | **Agree** | **Effective** |

To supplement the result presented in Table 2 are the actual scores obtained from the activities performed by the learner as shown in Figure 3. These activities are carried out by lesson levels with individual scores at each stage of the game such as at preparatory, developmental and evaluation stages.

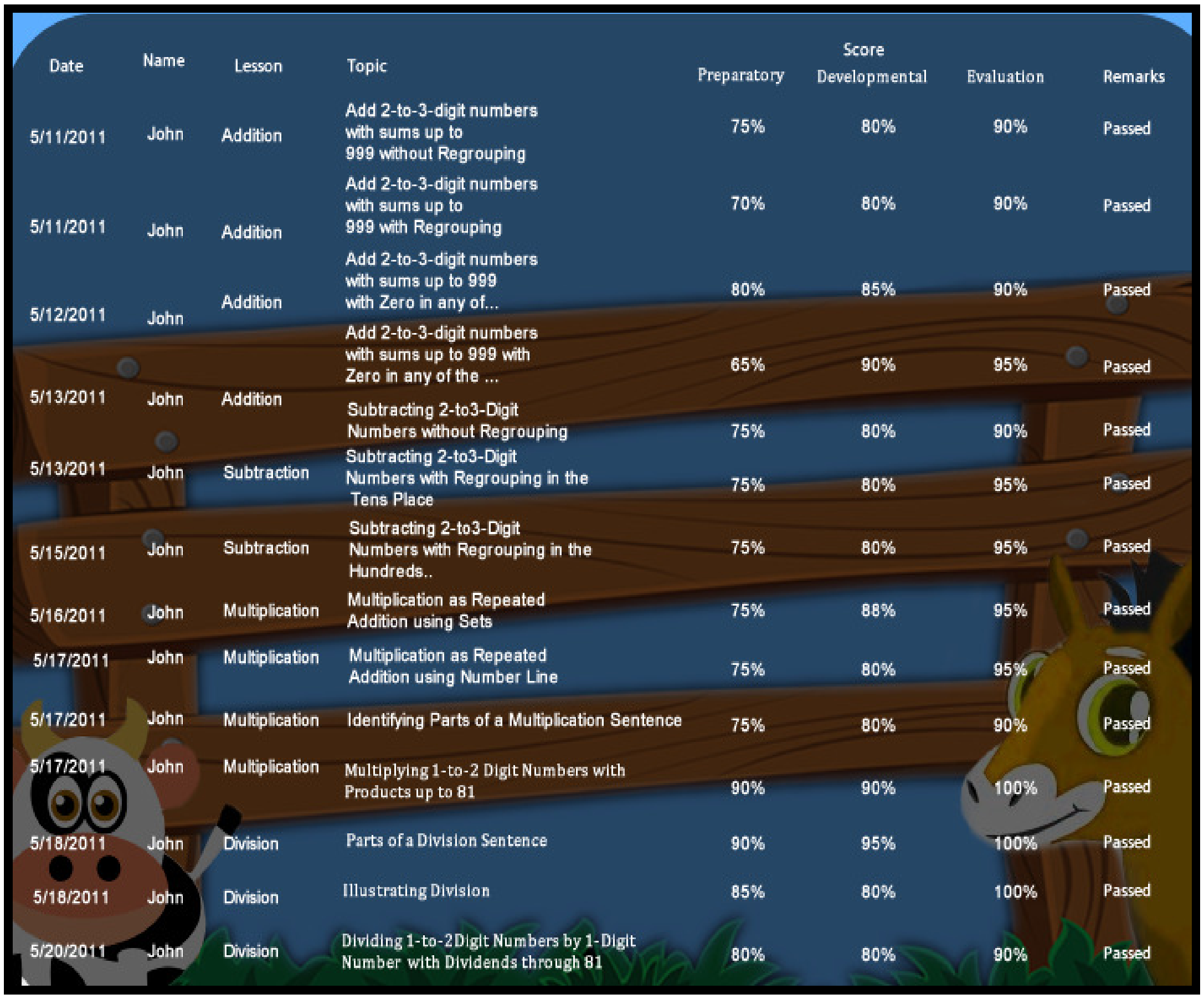

| Date | Name | Lesson | Topic | Score Preparatory | Score Developmental | Score Evaluation | Remarks |
|---|---|---|---|---|---|---|---|
| 5/11/2011 | John | Addition | Add 2-to-3-digit numbers with sums up to 999 without Regrouping | 75% | 80% | 90% | Passed |
| 5/11/2011 | John | Addition | Add 2-to-3-digit numbers with sums up to 999 with Regrouping | 70% | 80% | 90% | Passed |
| 5/12/2011 | John | Addition | Add 2-to-3-digit numbers with sums up to 999 with Zero in any of... | 80% | 85% | 90% | Passed |
| 5/13/2011 | John | Addition | Add 2-to-3-digit numbers with sums up to 999 with Zero in any of the ... | 65% | 90% | 95% | Passed |
| 5/13/2011 | John | Subtraction | Subtracting 2-to3-Digit Numbers without Regrouping | 75% | 80% | 90% | Passed |
| | | | Subtracting 2-to3-Digit Numbers with Regrouping in the Tens Place | 75% | 80% | 95% | Passed |
| 5/15/2011 | John | Subtraction | Subtracting 2-to3-Digit Numbers with Regrouping in the Hundreds.. | 75% | 80% | 95% | Passed |
| 5/16/2011 | John | Multiplication | Multiplication as Repeated Addition using Sets | 75% | 88% | 95% | Passed |
| 5/17/2011 | John | Multiplication | Multiplication as Repeated Addition using Number Line | 75% | 80% | 95% | Passed |
| 5/17/2011 | John | Multiplication | Identifying Parts of a Multiplication Sentence | 75% | 80% | 90% | Passed |
| 5/17/2011 | John | Multiplication | Multiplying 1-to-2 Digit Numbers with Products up to 81 | 90% | 90% | 100% | Passed |
| 5/18/2011 | John | Division | Parts of a Division Sentence | 90% | 95% | 100% | Passed |
| 5/18/2011 | John | Division | Illustrating Division | 85% | 80% | 100% | Passed |
| 5/20/2011 | John | Division | Dividing 1-to-2Digit Numbers by 1-Digit Number with Dividends through 81 | 80% | 80% | 90% | Passed |

Figure 3. Assessment Scores

Interestingness, enjoyment and engagement are likewise measured using the game playability indices in Table 3.

Table 3. Assessment on the Playability of Math World

| **Playability Attribute** | **Description** | **Mean** | **Interpretation** |
|---|---|---|---|
| Game Play | The game is enjoyable to replay | 3.47 | Moderately Agree |
| | The game drives you to play more rather than quitting | 3.50 | Moderately Agree |
| | The game provides clear goals throughout the play | 3.27 | Moderately Agree |
| | The games gives rewards that motivates the players to finish the game | 4.33 | Agree |
| | The player's fatigue is minimized by the game's different activities | 3.55 | Moderately Agree |
| | **Mean** | **3.62** | **Agree** |
| | | | |
| Game Story | The game has a single and consistent vision | 3.53 | Moderately Agree |
| | The game makes the players experience fairness of results | 3.33 | Moderately Agree |
| | The game makes the player think about the possible story outcome | 3.27 | Moderately Agree |
| | The game allows the player to use strategies while controlling his character | 3.34 | Moderately Agree |
| | The game brings the player into a level of personal involvement emotionally | 3.00 | Moderately Agree |
| | **Mean** | **3.29** | Moderately Agree |
| | | | |
| Mechanics | The players react according to players actions | 3.59 | Agree |
| | The player able to identify his score/status and goal in the game | 3.47 | Moderately Agree |
| | The game's control are consistent and easy to learn | 3.33 | Moderately Agree |
| | **Mean** | **3.46** | Moderately Agree |
| | | | |
| Usability | The player can easily turn the game on and off | 3.45 | Moderately Agree |
| | The player uses menus as part of the game | 3.49 | Moderately Agree |
| | The game has a stimulating sounds | 2.47 | Moderately Agree |
| | The game provides tips during the play | 3.23 | Moderately Agree |
| | The game intuitive and easy to learn menus | 3.43 | Moderately Agree |
| | The interface of the game is well-organized | 3.55 | Agree |
| | **Mean** | 3.27 | Moderately Agree |
| | | | |
| | **Overall Mean** | **3.41** | **Moderately Agree** |

# 6. CONCLUSION AND RECOMMENDATION

Many game-based VLEs have been developed recently for different purposes. In education, this is used as a pedagogical tool, allowing teachers and learners to have technologically-mediated activities in a fun way while maintaining the effectiveness of learning. In this paper, this is applied as supplement in teaching mathematics to second graders with the objectives to measure the learning effectiveness and playability of the game. With an aggregate of enhanced indices used to measure these educational attributes, results in assessing the effectiveness and playability of the game as well as assessment scores attest that objectives were met. This further implies that through VLE, learning mathematics is interesting and elicits motivation and enjoyment, hence a significant tool to complement the conventional approaches of teaching.

For future works, this paper recommends the environment to be enhanced, appropriate for multi-player and collaborative learning. Likewise, since mathematics is attributed to continuous practice and requires incremental learning through multi-variated simulations and exercises, more and other topics are encouraged to be included like those that require better comprehension for learners to acquire the skills for higher mathematics subjects .

# REFERENCES

[1] Barkand, J., Kush, & J. C., (2009) "GEARS a 3D virtual learning environment used in online secondary schools", Electronic Journal of e-Learning.

[2] Baxter, D. & Amory, A. (2004) "Development of a 3D Virtual Learning Environment to AddressMisconceptions in Genetics", In L. Cantoni & C. McLoughlin (Eds.), Proceedings of World Conference on Educational Multimedia, Hypermedia and Telecommunications 2004 (pp. 1256-1263). Chesapeake, VA: AACE

[3] Cook, A.D., (2009) "A case study of the manifestations and significance of social presence in a multi-learner virtual environment"

[4] Dalgarno, Barney, *The Potential of 3D Learning Environments : A Constructivist Analysis*

[5] Dalgarno, B. & Hedberg, J., (2000), "3D Learning Environments in Tertiary Education", Retrieved May 10th, 2011 from http://www.ascilite.org.au/conferences/melbourne01/pdf/papers/dalgarnob.pdf

[6] Dean, A. (2002). "Telelearning: invention, innovation, implications. Towards a manifesto", Electronic Journal of Instructional Science and Technology (in press).

[7] Dillenbourg, Pierre, (2000) "*Virtual Learning Environments*", University Of Geneva

[8] Driskell, J., (1992), Effect of over-learning on retention, Journal of Applied Psychology 77, 615-622.

[9] Eggan, P. & Kauchak, D., (2004) "Educational Psychology Windows on Classrooms"

[10] Frasca, G., (2004) " Videogames of the Oppressed: Critical thinking, Education, Tolerance, and other trivial issues"

[11] Ge, X., Yamashiro, K. A., a Lee, J., (2000) "Pre-class planning to scaffold students for online collaborative learning activities", Educational Technology Et Society

[12] Gros, B. (2007), *Digital Games in Education: The Design of Games-Based Learning Environments*

[13] Hogle, J., (1996), *Considering Games as Cognitive Tools: In Search of Effective Edutainment,* Retrieved from: ftp://twinpinefarm.com/pub/pdf/

[14] Hong, J., et al., *Assessing the Educational Values of Digital Games*

[15] Kiili, K. & Lainema, T., 2008. *Foundation of Measuring Engagement in Educational Games*. Jl. of Interactive Learning Research

[16] Lessig, Lawrence, (2000), *Code and Other Laws of Cyberspace*. Basic Books, ISBN 0-465-03913-8

[17] Nonis, D., (2005), *3D virtual learning environments (3D VLE)*, Ministry of Education, Singapore 2005. Retrieved April 4th, 2011 from: http://www.moe.gov.sg/edumall/rd/litreview/3d_vle.pdf

[18] Prensky, M., (2003), *Digital game-based learning.*

[19] Robertson & Howells, (2008), *Computer Game Design: Opportunities for successful learning*

[20] Squire, K., (2002), *Cultural Framing of Computer Video Games*, The International Journal of Computer Game research Volume 2, Issue 1

[21] Winn W., Hunter H., Hollander A., Osberg K., Rose H. & Char, P., (1997), *The Effect of Student Construction of Virtual Environments on the Performance of High-and-Low Ability Students*. A Paper presented at the Annual Meeting of the American Educational Research Association in Chicago. University of Washington.

[22] Yair, Y., Mintz, r. & Litvak, S., (2001), *3D-virtual reality in science education: An implication for astronomy teaching*. Journal of Computers in Mathematics and Science Teaching

[23] Trindade, Jorge, Fiolhais, Carlos, Teixeira, Victor Gil José, "Virtual Environment of Water Molecules for Learning and Teaching Science"

## Authors

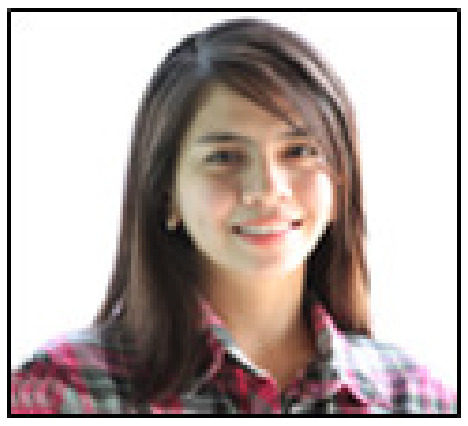

**Jean B. Maitem** graduated with a BS in Computer Science degree from Technological Institute of the Philippines (TIP) in 2008. She obtained her Master in Information Technology from the same school. She became a full-time faculty member of TIPQC CITE in June 2008.

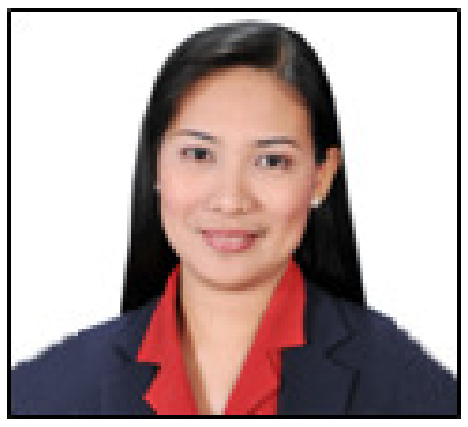

**Rosmina Joy M. Cabauatan** is an Assistant Professor of both Graduate School and College of Information Technology Education of Technological Institute of the Philippines. With her research interests in educational games, and applications of data mining in the computing and educational disciplines, she has been commended for his contributions to institutional researches. Pursuing her Doctoral Studies in Information Technology, she is currently undertaking her Dissertation on application of data mining in mobile computing.

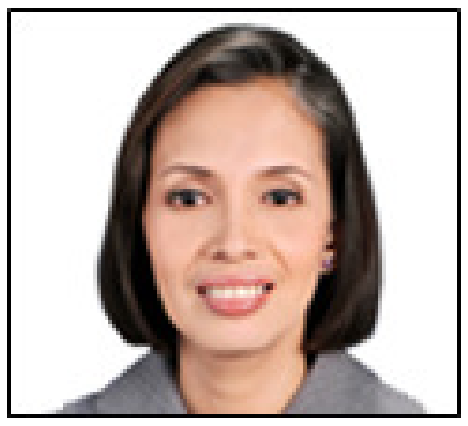

**Lorena W. Rabago** graduated with a BS Computer Science degree from the Philippine Christian University in 1990. She obtained her Master in Information Technology from Technological University of the Philippines. She is also a graduate of Doctor of Technology at the same university. Currently, the Department Head of the Information Technology in Technological Institute of the Philippines , a member of the Technical Committee on Information Systems of the Commission of Higher Education. She is also the Vice-President of the Philippine Society of IT Educators – National Capital Region Chapter and an Accreditor for Information Technology Education Faculty, Instruction and Laboratories of the Philippine Association of Colleges and Universities of the Commission on Accreditation.

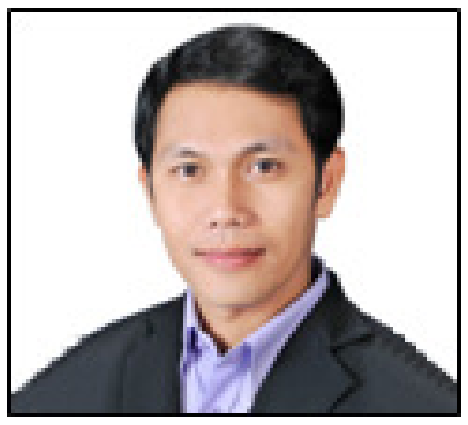

**Bartolome T. Tanguilig III** graduated with a BS in Computer Engineering degree from Pamantasan ng Lungsod ng Maynila in 1991. He obtained his Master in Computer Science from De La Salle University in 1999. He is also a graduate of Doctor of Philosophy in Technology Management from Technological University of the Philippines in 2003. He is the founder of the Junior Philippine ITE Researchers. Currently, the Dean of the College of Information Technology Education in Technological Institute of the Philippines, a member of the Technical Panel in Information Technology Education and the Chair of the Technical Committee on Information Technology of the Commission on Higher Education, national board member of the Philippine Society of IT Educators. He became an accreditor of Graduate Programs of the Philippine Association of Colleges and Universities of the Commission on Accreditation in 2010.